# The analysis of predictability of α-decay half-life formulae and the α partial half-lives of some exotic nuclei.

(Short title: Analysing the predictability of α-decay half-life formulae)

N. Dasgupta-Schubert [1,2]*, M.A. Reyes [3] and V.A. Tamez [4]

[1] Inst. Invest. Químico-Biológicas, [2] Lab. Radiación, Inst. Física y Matemáticas,
Univ. Michoacana, Morelia, Mich. C.P. 58060, México
[3] Dept. Elect. Eng., Univ. of Texas Pan American, Edinburg, TX, 78539, USA
[4] Dept. Chem., Texas A&M Univ., Kingsville, TX 78363, USA

*email: nita@ifm.umich.mx

**Abstract**: The predictabilities of the three α-decay half-life formulae, the Royer GLDM, the Viola-Seaborg and the Sobiczewski-Parkhomenko formulae, have been evaluated by developing a method based on the *ansatz* of standard experimental benchmarking. The coefficients of each formula were re-derived using the reliable data of the α-standards nuclei. The modified formulae that resulted were used to evaluate the accuracies of the formulae towards the prediction of half-lives of a set of nuclides with well-studied α-spectroscopic data as well as a set of exotic α-emitters. Further, a simple linear optimization of the modified formulae allowed adjustments for the insufficient statistics of the primary data set without changing the modified formulae. While the three modified formulae showed equivalent results for all the medium heavy nuclei except the odd-odd, the modified GLDM showed relatively the best figures of merit for the odd-odd and superheavy nuclides.

1.  **Introduction**

Alpha decay is one of the two main decay modes of the heaviest nuclei, termed Super-Heavy Elements (SHE), and constitutes one of the dominant decay modes of highly neutron deficient medium mass nuclei [1,2]. In hot-fusion heavy ion reactions, product nuclides closest to the shell model predicted *Island of Stability*, decay by the emission of *α* particles and successive product nuclei are genetically linked by the chain of *α* decays [1, 3, 4]. Thus identifying and characterizing the *α* decay chains form a crucial part of nuclide identification in the synthesis of SHE [1, 4, 5]. The experimental identification is aided by theoretical predictions of $\alpha$ decay half-lives and decay energies. Estimates of decay half-lives are used to corroborate the identification of short-lived isotopes by recoil separation techniques (e.g. as described in [1]); for the nuclear chemist, an *a priori* knowledge of the half-lives of the longer-lived product nuclides in heavy ion fusion reactions is important for the design and execution of experiments for the characterization of the radiochemical properties of the heavy element. Radiochemistry often serves as a parallel investigation of product nuclei in SHE synthesis [1, 5, 6].

The $\alpha$ decay mechanism is theoretically described by the quantum mechanical tunneling through the potential energy (PE) barrier leading from the mother nucleus´s energy to the total energy of the daughter nucleus and $\alpha$ particle. The net energetics of the process manifest as the alpha decay Q value, $Q_\alpha$. The PE barrier penetration probability is deduced from calculations such as the WKB, leading to the $\alpha$ decay constant, $\lambda_\alpha$ that is related to the $\alpha$ decay half-life, $T_\alpha$, as $\lambda_\alpha = \ln2/T_\alpha$. Consequently, the predicted half-life remains very sensitive to the shape and energetics of the barrier, which as such, serves to test the particular theoretical model of the potential energy surface of the decaying nucleus. The discovery of new $\alpha$ emitters have spurred theoretical studies on $\alpha$ decay; for example within the relativistic mean field theory [7], the DDM3Y interaction [8, 9], the generalized liquid drop model (GLDM) [10], the Skyrme-Hartree-Fock mean-field model [11] and the analytical superasymmetric fission model [12]. Some of the theoretical calculations, *e.g.* the GLDM, as well as some phenomenological observations have been reduced to analytical formulae that connect the $\log(T_\alpha)$ with the $Q_\alpha$ and the (Z,A) of the parent nuclide and wherein the coefficients are typically obtained from fits to known experimental half-lives, $T_\alpha^{exp}$ [10, 13-17, 19]. The analytical formula is of practical value for the rapid prediction of $\alpha$ half-lives, for the observation of systematics over a wide (Z,A) region and because an expansion/refinement of the experimental database permits a relatively facile upgrade of its coefficients. It certainly cannot replace a deep theoretical study of $\alpha$ decay which aims at a microscopic description of the underlying mechanism. As mentioned above, in the experimental synthesis of SHE, analytical formulae have been used for the additional verification of the nuclidic assignment, *e.g.* the Viola-Seaborg formula by the Dubna group [1, 18], the Sobiczewski-Parkhomenko formula by the Dubna-Darmstadt group [6]. Therefore it becomes useful to analyse the predictive accuracy of several well-known and recent analytical formulae in a rigorous manner to better understand their reliability. To date, this has received insufficient attention.

The degree of closeness between the formula predicted ($T_\alpha^{calc}$) and the $T_\alpha^{exp}$ of newly discovered α emitters has generally been taken as a measure of the goodness of the formula towards the prediction of unknown half-lives [9, 10, 14, 19]. Such an accuracy test is compromised by the substantial uncertainties in the experimental α spectroscopic and mass data of the exotic alpha emitters - the number of those with well-characterized data are too few to permit a sufficiently conclusive statistical evaluation. Additionally, the coefficients of the formula are obtained by a fitting procedure that involves the experimental data of a large number of α emitters (the primary basis set). While being statistically beneficial, this inevitably causes the inability to allocate a large set of well-characterised α emitters for the independent test of accuracy of the formula (because the full set of well-known α emitters are already included in the primary set). The lack of consensus as to what constitutes an acceptable primary set could possibly lead to inadvertent oversights. In [19] a primary data set that was slated towards the heavy nuclides so as to overlap the (Z,A) range of the SHE for which the predicted half-lives were tested, might have resulted in a 'cyclic argument', *i.e.* the double inclusion of some of the nuclides in both the coefficient-determining and the accuracy-testing sets. Furthermore, a wide (Z,A) spanning range of the primary data set increases the likelihood of the inclusion of nuclides, especially at the extrema, whose decay data are uncertain – most often poorly determined α branching fractions (*e.g.* $^{110,111}$Xe, $^{111}$I and $^{189}$Bi) or, in the case of many of the heavy nuclides with Z>100, $Q_\alpha$s that are surmised from systematics only [20]. This inclusion causes slight variations in the values of the coefficients whose effects on the calculated half-lives are amplified because of the logarithmic relationship. In a previous work [21], we had devised a method to examine in detail, the predictability of the GLDM formula. In this work, the method is extended to cover a systematic investigation of the reliability of the three main extant analytical formulae of alpha decay half-lives: the Sobiczewski modified semi-empirical Viola-Seaborg formula [16], the recent phenomenological formula of Sobiczewski and Parkhomenko [17] as well as the GLDM based formula of Royer [10]. The *ansatz* for this stringent method of evaluation and inter-comparison is borrowed from standard benchmarking procedures in experiment. Concomitantly, we check the utility of the modified formulae with revised coefficients that result using this method, for their own predictive accuracies for the SHE and other newly discovered alpha emitters.

## 2. Alpha Decay Formulae

*The Royer Formula (R)*

The GLDM including proximity effects between the α particle and the daughter nucleus, is utilized in the description of the PE barrier and is adjusted to reproduce the experimental $Q_\alpha$; the $T_\alpha$ is deduced from the WKB barrier penetration probability as for spontaneous asymmetric fission. The model expressions involving Z,A and $Q_\alpha$ were fitted against the $T_\alpha^{exp}$, for 373 α emitters [10] with Z and A in the range of 52-111 and 107-272 respectively, to arrive at the 12-parameter analytical formula,

$$\log_{10}[T_\alpha(R)] = a + b.A^{1/6}.Z^{1/2} + c.Z/(Q_\alpha)^{1/2} \quad (1)$$

where $T_\alpha(R)$ is the calculated alpha partial half-life in seconds and $Q_\alpha$ is in MeV. The values of the coefficients a, b and c for parent nuclides of Z proton-N neutron even-even (e-e), even-odd (e-o), odd-even (o-e) and odd-odd (o-o) respectively were, (-25.31, -1.1629, 1.5864); (-26.65, -1.0859, 1.5848); (-25.68, -1.1423, 1.592); and (-29.48, -1.113, 1.6971).

*The Viola-Seaborg-Sobiczewski Formula (VSS)*

In 1966, Viola and Seaborg, generalised the empirical Geiger-Nuttall formula [13], to include more adjustable parameters and obtained the following 7-parameter formula [15] for the calculated alpha partial half-life $T_\alpha(VSS)$ in seconds ,

$$\log_{10}[T_\alpha(VSS)] = (a.Z + b).(Q_\alpha)^{-1/2} + (c.Z + d) + h_{log} \quad (2)$$

where the coefficients a, b and c were obtained from fits to e-e α decaying nuclei and $Q_\alpha$ is in MeV. The parameter $h_{log}$ is the hindrance factor for nuclei with unpaired nucleons and is obtained by fits to odd nuclei. Its value for e-e, e-o, o-e and o-o Z-N numbers of the parent nuclide were obtained respectively as, 0, 1.066, 0.772, 1.114. For the a, b, c and d parameters we consider the Sobiczewski modified values obtained using the more recent and expanded data base of e-e nuclides [16], which are: a = 1.66175; b = -8.5166; c = -0.20228; d = -33.9069

*The Sobiczewski-Parkhomenko Formula (SP)*

Recently Sobiczewski and Parkhomenko have introduced a 5-parameter phenomenological formula for the alpha partial half-life $T_\alpha(SP)$ in seconds [3, 17] motivated by the need to simplify the VSS formula as follows,

$$\log_{10}[T_\alpha(SP)] = a.Z.(Q_\alpha - \bar{E}_i)^{-1/2} + b.Z + c \qquad (3)$$

where $Q_\alpha$ is in MeV. The parameters, a, b and c are respectively: 1.5372; -0.1607; -36.573. The parameter $\bar{E}_i$, the average excitation energy of a state of the daughter nucleus to which the α decay goes, is given by,

$\bar{E}_i = 0$ for e-e; $\bar{E}_i = \bar{E}_n = 0.171$ MeV for e-o; $\bar{E}_i = \bar{E}_p = 0.113$ MeV for o-e; $\bar{E}_i = \bar{E}_p + \bar{E}_n$ for o-o.

The values for the 5 parameters above have been obtained by the adjusting eqn. 3 to the experimental $T_\alpha$ and $Q_\alpha$ of 201 α decaying nuclei between Z=84-111 and N= 128-161. However, nuclei with N close to the shell closures at N=152 and 162 were omitted.

The large sizes of the primary basis sets used to arrive at the coefficients for all three formulae are beneficial for the statistics but their variability in both size and in the (Z,A) identity of the nuclides vitiates a meaningful inter-comparison of accuracy. Additionally, all the considerations mentioned in the Introduction, apply.

3. **Method**

The method is based on the *ansatz* of standard bench-marking procedures in experiment: a fixed-size basis set of nuclides with very well-characterised alpha spectroscopic and mass data are chosen to derive anew the coefficients of each of the three formulae, which are now the 'calibrated formulae'. This set of reference nuclei, termed REF nuclei, are the alpha energy and intensity standards used in alpha spectrometry ("Alpha Particle Energy and Intensity Standards", p.14064 of [20]): 78 nuclides between the Z of 62 to 100 and the N of 84 to 155. This is not a very large basis set with wide Z,N spanning members so that the resulting formulae (all with the suffix *m*, see below and tables 1b - 4) are not expected to produce a close coincidence between $T_\alpha^{calc}$ and $T_\alpha^{exp}$ for nuclei far outside this range. However, that is not the main intention of the present work. The intention is to produce formulae with highly reliable coefficients that can be used to test the predictability of the formulae using an independent and approximately similar data set. The secondary question of how good the calibrated formulae are with respect to the prediction of the α partial half-lives of the exotic α emitters and the effect of the REF basis set statistics on these, is treated in section 4.4.

In this work the term *predictability* is used to connote the extent of accuracy of the three formulae in section 2 towards the prediction of the half-lives of an independent set of nuclei, i.e. nuclei not used in coefficient-determining set. This test of accuracy necessitates the experimental nuclear data of this independent set of nuclides, the 'TEST' data set, to be also well known. The TEST data set was chosen (criteria presented in [21]) to consist of experimentally well-studied α-decaying nuclides [20], 235 in number, that cover a (Z,A) range close to the REF nuclei in order to reduce the possibility of the influence of Z,A dependent differences in the underlying nuclear parameters in the data analysis. Thus, the general equivalence in the type of nuclides and quality of experimental data between the REF and TEST sets are ensured. The details of how the experimental data for the REF and the TEST were extracted as well as their full listing, are provided in reference [21].

The coefficient of variation CV of the experimental quantity of interest is defined as the ratio of the standard deviation of the experimental quantity to its average value

[23]. Table 1a shows the root mean square values of the CV for $T_\alpha^{exp}$ and $Q_\alpha^{exp}$, (RMSCVT$_\alpha^{exp}$ and RMSCVQ$_\alpha^{exp}$), for the REF and TEST nuclides. The rather low values for the REF for all nucleon parity sets substantiate their status as $\alpha$ energy and intensity standards. The values for the TEST are also low which justifies their use as tool for bench-marking the formulae. Also, errors on the $Q_\alpha^{exp}$ of both are practically negligible.

The REF and TEST sets are the analogues of the 'calibration standard' and 'standard reference material (SRM)' in experimental practice where the latter is the well-known probe used to evaluate the experimental method/instrument after its calibration. In contrast, the precision errors on the experimental values of the exotic nuclei – the SHE and medium-mass $\alpha$-decaying nuclides - considered in the present work and shown in table 3, are large. This set has been upgraded from the one considered earlier in [21] in view of the latest experimental results [22, 24]. For many, the $\alpha$ branching fractions and the standard deviations on the $Q_\alpha^{exp}$ values are not known or not quoted in the standard nuclear data libraries or the $T_\alpha$ have not been sufficiently corroborated by independent groups, for which reason the RMSCVT$_\alpha^{exp}$ and RMSCVQ$_\alpha^{exp}$ have not been computed. Thus the efficacy of these exotic nuclei as instruments of evaluation of the formula´s predictability [*cf.* 10, 15, 17 and 19] is compromised by the inherent non-negligible errors in the data values.

3.1 *The Modified Formulae*

The standard Levenberg-Marquardt non-linear multivariate regression fitting algorithm [25] was used to derive the coefficient values of the formulae and their errors using the experimental $\alpha$ half-lives and errors of the REF basis set. These resulted in the `calibrated´ formulae whose coefficients are shown in table 1b. Rm stands for the modified Royer formula.

In the case of the Viola-Seaborg (VS) formula, two approaches were used to derive the coefficients of the modified formula, resulting in the formulae VSm1 and VSm2. In the former, only the e-e REF basis was used to derive the a,b,c,d coefficients of eqn. 2 and the odd (Z,N) bases to derive the $h_{log}$ parameter (as per the original authors). Thus VSm1 also is a 7 parameter formula. In the case of VSm2, $h_{log}$ was merged into the parameter d and all the 4 coefficients a to d were re-derived separately using the experimental data for all the four parity sets of REF, resulting in a 16 parameter formula.

In the case of the Sobiczewski-Parkhomenko (SP) formula, the same two approaches were used to obtain the modified formulae, SPm1 and SPm2. For the former, the e-e REF data set was used to derive the coefficients a,b, c and the odd-nucleon sets were used to derive the values of $E_i$, resulting as before, in a 5 parameter formula. For the SPm2 formula, all the four parity sets of REF were used to derive a,b,c and $E_i$ separately resulting in a 15 parameter formula.

Consistent with previous practice [21, 10, 22] the figure of merit (FOM) used to evaluate the various modified formulae is the standard criterion [23, 26]: the relative deviation or relative error RE between the experimental and calculated $\log_{10}T_\alpha$,

$$RE = (\log_{10}T_\alpha^{exp} - \log_{10}T_\alpha^{calc}) / \log_{10}T_\alpha^{exp} \qquad (4)$$

Additionally, following Sobiczewski and Parkhomenko, we determine the values of the index $\bar{f}$, derived from their definition [17] of the average value $\bar{\delta}$ of the discrepancy δ,

$$\bar{\delta} = \frac{1}{N}\sum_{i=1}^{N} \left|\log_{10}\left(T_{\alpha i}^{calc}/T_{\alpha i}^{exp}\right)\right| \qquad (5)$$

$$\bar{f} = 10^{\bar{\delta}} \qquad (6)$$

The features of $\bar{f}$ are discussed in [3]. Although it is a more algebraically 'processed' FOM (whereas the RE relates directly to the calculated and measured quantities), it has been included in the evaluation of the TEST because it provides a means of estimating the average multiplicative factor of $T_\alpha^{calc}$ over $T_\alpha^{exp}$.

The comparison with results using the original formulae are meaningful only in the context of the exotic nuclei because the coefficient-deriving (or 'calibration') basis sets of the originals differ significantly from REF and include both REF and TEST as sub-sets.

4. **Results and Discussion**

### 4.1 *The calibration nuclide set: REF*

Table 1b shows the root mean square values of the RE, the RMSRE, for the different modified formulae. The RMSRE (or a similar index) when deduced for the 'calibration' set is a measure of the accuracy or goodness of fit of the formula with the experimental data. This, and observations on predictability based on the same index for exotic nuclei with (Z,A) slightly beyond the calibration set, are essentially what previous works [10, 15-17] have done, albeit some like Sobiczewski and Parkhomenko [17] have discussed the conjoint sets in slightly greater detail, graphically. The difference in this work lies in the distinction made between testing the goodness of fit (REF), testing the predictability using a viable and independent data set (TEST) and finally using the results for prognostics on a frontier set (the exotics).

Comparing between the RMSRE values for the m1 and m2 variants of the VS and SP formulae, table 1b shows that for all parity sets the fits are better for the m2 formulae. Infact taking the average RMSRE over all parity sets, the values for VSm1, VSm2, SPm1 and SPm2 are 0.1070, 0.0793, 0.1151 and 0.0784 indicating a 34.9% and 46.8% improvement in quality of overall fit of the VSm2 over the VSm1 and the SPm2 over the SPm1 formulae respectively. Given the negligible random errors of the experimental data of REF, these differences could be meaningful. Their significance is discussed in greater detail in section 4.2.

Amongst the Rm, Vsm2 and SPm2 formulae, the lowest RMSRE for all parities except the o-o are for the Rm. The values of the average RMSRE for these sets for the Rm, VSm2 and SPm2 formulae respectively are 0.0571, 0.0586 and 0.0586 – very close and indicating an almost equivalently good fit for all three formulae. However, the poor statistics of the o-o REF set causes greater fluctuations resulting in higher and more disparate values of the RMSRE amongst which the lowest is for the SPm2 and the relative highest for the Rm. Thus we would expect greater anomalies in the predictability of these formulae when applied to the o-o sets such as those for TEST and the exotics.

### 4.2 *The probe nuclide set: TEST*

The various modified formulae (section 3 and table 1b) were applied to the 235 nuclides of the TEST set. The standard deviations s(RE) on the RE values were obtained by the propagation of the experimental errors and the errors in the formula over the RE expression (eqn. 4). The RMSRE along with its corresponding RMS value of the standard deviation, RMS[s(RE)], is shown in table 2. The RMSRE±RMS[s(RE)] yields a measure of the accuracy of the various formulae in their prediction of the $\alpha$ half-life and the uncertainty of the predictability associated with the dispersion caused by the experimental and formula precision errors. The results for each formula are summarised for the 4 parity sets by the average RMSRE±RMS[s(RE)]. Between the m1 and m2 variants, the m2 formulae have the lower values of the FOM. To test whether the differences are significant or not the Student's t test [23, 26] was carried out for the odd parity sets for these formulae using the RMSRE±RMS[s(RE)] (for the e-e case, the m1 and m2 formulae are the same). The computed and tabulated t values for the formula pairs are shown at the bottom of table 2. Within a 95% and/or 90% confidence limit, the t test results in the differences being significant which indicates a better degree of predictability of the m2 formulae. This result is not unexpected in view of the fewer number of adjustable parameters of the m1 formulae. Possibly, the nature of the PE barrier may imply the need to include a larger number of parameters, in a sense supporting the validity of the m2 formulae.

Within the Rm, VSm2 and SPm2 formulae, the average values of the RMSRE show that the 3 formulae are virtually the same in accuracy with the slight advantage to VSm2. A parallel comparison with the average values of $\overline{f}$ (not a standard FOM) show that the Rm and VSm2 are again close with the former now as the relative best and the SPm2 the worst. This apparent difference is the consequence of the inhomogeneity in the distribution of the discrepancies for the large TEST set (a similar observation was made in [17]).

The high value of both FOMs for the o-o set in SPm2 could likely be the result of the poorer o-o REF set statistics. The SPm2 formula for the o-o set requires 4 adjustable parameters to be computed in a relatively more complex expression of the term containing the dominant variable in the formula, the $Q_\alpha$, so that poorer statistics could probably cause greater fluctuations. The VSm2 formula also requires 4 parameters to be computed but its simpler form probably contributes to its better FOM. Relatively the best FOM is for the Rm formula which requires the computation of only 3 parameters in a formula that is functionally simple in its $Q_\alpha$ term. Additionally it includes an $A^{1/6}$ dependence that is absent in the other two formulae. Sobiczewski and Parkhomenko discuss [17] the modification of the $\bar{E}_i$ term in their formula to include a $A^{1/3}$ term that would take account of (within a simple model of the harmonic oscillator) the density of single-particle levels increasing with A and conclude that it did not improve the description of the $T_\alpha$ of the heaviest nuclei. The $(ZA^{1/3})^{1/2}$ term in Royer's formula however, may contribute to the improvement of predictability especially for higher (Z.A) nuclides, which we test further in the section 4.3 on the application of the modified formulae to the SHE.

The majority of the RE and s(RE) values for the TEST nuclides are quite small but there are marked deviations for certain nuclides that result in the relatively high RMSRE±RMS[s(RE)] values, as for example for the e-e set for all formulae and for the relatively high value for the e-o set for Rm. These high values occur mostly at certain N numbers. Unlike [17] no attempt has been made in this work to selectively exclude nuclei near neutron shell closures. Only in the case of the SPm1 for the e-o set, the very large value of s(RE) for $^{215}_{90}Th_{125}$ led to its exclusion and it was also dropped for SPm2 for the sake of consistency. Fig. 1 shows the distribution of RE with respect to N for all nuclides in TEST and for all formulae. Marked non-zero magnitudes of the RE occur in regions of N near the shell closure numbers of 82, 126 and N=146 ([27]: 'deformed magic') as well as around the regions of N=88-90 ([28]: 'shape transition'), N=96-98 ([29]: 'loss of collectivity and sub-shell gaps') and N=134 ([30]: 'static octupole deformation').

4.3  *Linear optimization*

Since the small statistics of the odd parity sets of REF, particularly for the o-o set, cause a statistical disadvantage to the coefficients of the modified formulae, we attempt to *pseudo* increase the statistics without deviating from the basic premise, i.e. the derivation of coefficients using a fixed highly reliable basis set so that the coefficients are not subject to arbitrary changes. Accordingly, the log of the calculated $T_\alpha$ values, $\log_{10}(T_\alpha$ calc) obtained from the modified formulae for the TEST odd-nuclides were linearly fitted to their $\log_{10}(T_\alpha^{exp})$ values. The resulting linear regression equation, $\log_{10}(T_\alpha^{lin-opt})$ was taken to be the 'statistics-wise improved' version of the corresponding modified formula. In the present work, this technique first applied to the Rm formula in [21] is now extended to all three.

$$\log_{10}(T_\alpha^{lin-opt}) = \mu \log_{10}(T_\alpha^{calc}) + \kappa \qquad (7)$$

The values of $\mu$ and $\kappa$ for the e-o, o-e and o-o sets of the Rm formula are respectively, {0.9913, -0.2811}, {1.0431, 0.3491}, {0.9093, 0.5284}. For the VSm2 formula they are respectively, {0.9943, -0.1805}, {1.0555, 0.3363}, {0.8973, 0.4120}; and for the SPm2 formula they are respectively, {0.9988, -0.2400}, {1.0558, 0.3631}, {0.7489, 1.1479}. The linearly optimized formulae can only be used on a set of $\alpha$-emitters outside the REF and TEST.

### 4.4    The application nuclide set: EXOTICS

Having analysed the predictability of the three modified formulae it would be interesting to see how they, as well as their linearly optimized variants, compare with the original formulae for the computation of the $T_\alpha$ of the SHE and medium mass $\alpha$-emitters (the EXOTICS). Table 3 shows the $\log_{10}T_\alpha^{calc}$ values for the three modified formulae along with their calculated standard deviations (only the m2 formulae have been chosen in view of the aforesaid results) and table 4 shows the FOMs of the original and the modified formulae as well as the linearly optimized modified formulae for the odd sets. It can be seen from table 3 that the agreement between the experimental and calculated $\log_{10}T_\alpha$ and between the calculated values themselves are generally the best at the centre of the range. The precision errors in the calculated values do not include the $Q_\alpha^{exp}$ errors because of insufficient experimental data. As example however, a typical 40 keV FWHM uncertainty in the decay energy is seen to cause a fair increase of the precision error of $\log_{10}T_\alpha^{calc}$ for the particular nuclides considered..

Comparing between the original formulae only (table 4), the values of the average

FOMs show the R formula to be the best followed by the SP and then very closely by the VSS. The position of the VSS is understandable from the fact that the coefficients of the R and SP formulae have been recently derived using large calibration sets with a few members that overlap the exotic α-emitters considered in table 3 (both extend the Z range to 111) whereas the VSS coefficients are older, derived from a less (Z,A)-extensive data base. It is this overlap as well as the much larger statistics of their calibration sets that account for the better FOMs of the original formulae over the modified formulae. This is particularly evident for the o-o set. Table 4 shows that the ratios of the average FOMs of the modified formulae to the originals are 1.59, 3.18 and 4.45 for the Rm, VSm2 and SPm2 respectively, the >1 values driven largely by the high FOMs of the o-o set. If the o-o set is excluded the ratios more closely approach 1, being 1.24, 1.03 and 1.05 respectively. This shows that despite the much narrower (Z,A) range and the fewer overall statistics of the calibration sets of REF and the fact that nuclides near neutron shell closures were not excluded, the modified formulae do rather well.

Comparing between the set of modified non-linearly optimized formulae alone, table 4 shows that the best average FOM (1.21) is for the Rm followed by the VSm2 (3.33) and finally by the SPm2 (4.61). This more or less follows the same trend observed for the TEST set, albeit here the differences are wider. In all three, the e-e set possesses comparatively the better FOMs, attributable to the higher statistics of the REF calibration set (although the presence of some exotics with RE spikes (discussed later) causes random increases in the RMSRE). The effect of excluding the e-e set on the average RMSRE (last column of table 4), results in increased values with the same trend as before. From the values of the FOMs for the linearly optimized modified formulae in table 4, we note that the overall deviations are significantly reduced. The most noticeable is the reduction in RMSRE for the o-o set: the ratios of the FOMs of the lin-opt formulae to the modified formulae decrease as 0.48, 0.81 and 0.56. If the RMSRE for the e-e sets of the Rm, VSm2, SPm2 formulae are combined with the RMSRE of the odd sets of the corresponding linearly optimized modified formulae, the resulting average RMSRE are now 0.86 (Rm&lin-opt-Rm), 2.76 (VSm2&lin-opt-VSm2) and 2.94 (SPm2&lin-opt-SPm2). These are substantial reductions. In sum, for the exotic nuclei, the combined Rm perform the best followed by the nearly equivalent performance of the combined VSm2 and the combined SPm2.

The aforesaid suggests that this technique of linear optimization that 'augments' the statistics of the calibration set to improve the predictability of the formula may be a useful alternative. Its chief advantage lies in the fact that the coefficients of the formula calculated using a highly reliable primary data set (the calibration set), do not have to be re-calculated. Rather the intact formula has only to be locally adjusted using simple linear regression. The larger the number of nuclei found with the same quality of data as TEST near the zone of the unknown nuclei, the more accurate the linear optimization is likely to be.

The variation of the RE for the different formulae with respect to the N of the exotic set are plotted in fig.2. For most nuclei the RE are close to zero, but for some, especially those near the shell closures of N=126 and around N=152, 162 and 172 [31], there are large deviations. Linear optimization reduces these deviations for all formulae but their signatures persist. All three formulae do not invoke a direct dependence on the shell structure - only an indirect one (via the $Q_\alpha$).

## 5  Conclusion

The predictabilities of the three commonly used alpha decay half-life formulae, the Royer GLDM (R), the Viola-Seaborg (VS) and the Sobiczewski-Parkhomenko (SP) formulae, have been evaluated in a careful way by extending a method developed by us earlier [21]. The coefficients of each formula are derived using the highly reliable data of the α-standards nuclei (the calibration set: 'REF') resulting in the modified formulae, Rm, VSm1, VSm2, SPm1 and SPm2, where the distinction between the m1 and m2 is that in the latter all four parameters are free and evaluated independently. The m2 variants are shown to yield a better figure of merit (FOM). The goodness of fit of the functions of the three formulae, Rm, VSm2 and SPm2 are very close, with the SPm2 as the relative best. The modified formulae are used to benchmark the accuracies of the formulae towards the prediction of half-lives of a set of nuclides with well-studied α-spectroscopic data (the probe set: 'TEST'). The predictabilities of the three formulae are close except when the REF's statistics are poor, e.g. the o-o REF set. In such a case the

Rm emerges as the relative best. The efficacy of the formulae outside the (Z,A) region of the REF and TEST, is tested on the SHE and newly discovered α-decaying nuclides (the application set: 'EXOTICS'). Additionally, a simple linear optimization of the modified formulae allows adjustments for insufficient statistics of the REF data set without changing the modified formulae. The Rm along with the linearly optimized Rm, perform the best for this set. The better performance of the GLDM formulation for heavy nuclides perhaps may be due to its inclusion of the $(ZA^{1/3})^{1/2}$ term which distinguishes it from other phenomenological formulae based on the Geiger-Nutall observation. However, since all the formulations do not directly include shell-structure effects, large relative deviations show up in the region of neutron shell closures and related structures. The poor statistics of the odd REF sub-sets points to a need by nuclear data groups to study and include more nuclides as standards in this category. Finally, the method described in this work is general and robust (it employs only the α-standards), and can be extended to other less-common or newer analytical alpha half-life formulae.

**Acknowledgement**: NDS expresses her gratitude to Prof. Sudip K. Ghosh of S.I.N.P Kolkata, and to the referees for their critiques, and to Profs. A. Türler, A. Sobiczewski and Y. Oganessian for helpful discussions during the ENAM08 conference. This work was partially funded by the CIC of the Universidad Michoacana (NDS) and by the LSAMP scholarship of the University of Texas (MAR, VAT).

Table 1a: The root mean square values of the coefficients of variation (CV), $RMSCVT_\alpha^{exp}$ and $RMSCVQ_\alpha^{exp}$, for the experimental α half-lives and $Q_\alpha^{exp}$ values [21] for the REF and TEST nuclides of even or odd proton (Z) and neutron (N) numbers. The 78 REF are the α-standards nuclides and the 235 TEST are well-studied α-decaying nuclei. The standard deviations of the $T_\alpha^{exp}$ values used for the calculation of the CV, were obtained by propagation of the errors on the total half-lives and α-branching fractions of the nuclides [21].

| Basis set | Z,N | n | RMSCV($T_\alpha^{exp}$) | RMSCV($Q_\alpha^{exp}$) |
|---|---|---|---|---|
| REF | e-e | 31 | 9.143x10$^{-2}$ | 4.658x10$^{-4}$ |
|  | e-o | 19 | 7.124x10$^{-2}$ | 3.900x10$^{-4}$ |
|  | o-e | 15 | 1.146x10$^{-1}$ | 4.535x10$^{-4}$ |
|  | o-o | 13 | 1.464x10$^{-1}$ | 5.640x10$^{-3}$ |
| TEST | e-e | 78 | 1.818x10$^{-1}$ | 1.101x10$^{-3}$ |
|  | e-o | 66 | 2.338x10$^{-1}$ | 5.119x10$^{-3}$ |
|  | o-e | 55 | 2.312x10$^{-1}$ | 4.419x10$^{-3}$ |
|  | o-o | 36 | 2.154x10$^{-1}$ | 4.750x10$^{-3}$ |

Table 1b: The values of the coefficients of the modified Royer GLDM (*Rm*), the modified Viola-Seaborg (*VSm1* & *VSm2*) and the modified Sobiczewski-Parkhomenko (*SPm1* & *SPm2*) formulae obtained using the REF nuclide basis set. For the e-e sub-set, the *m1* and *m2* formulae are the same. The column *RMSRE* is the root mean square of the relative error between the logarithms (base 10) of the experimental and calculated α half-lives for the REF set.

| Formula | Z,N | Coefficients | | | | | | RMSRE |
|---|---|---|---|---|---|---|---|---|
| | | a | b | c | d | $h_{log}$ | $E_i$ | |
| Rm | e-e | -22.2505 ±0.2465 | -1.191 ±0.013 | 1.5226 ±0.0018 | - | - | - | 0.0434 |
| | e-o | -27.773 ±0.2833 | -1.0032 ±0.0045 | 1.5809 ±0.0041 | - | - | - | 0.0733 |
| | o-e | -29.2562 ±0.4892 | -1.0279 ±0.0077 | 1.6198 ±0.0070 | - | - | - | 0.0546 |
| | o-o | -28.3003 ±0.0528 | -1.3553 ±0.0161 | 1.8264 ±0.0101 | - | - | - | 0.1899 |
| VSm1 | e-e | 1.4833 ±0.0115 | 2.0678 ±1.1487 | -0.2107 ±0.0045 | -30.533 ±0.4533 | 0.0 | - | 0.0469 |
| | e-o | " | " | " | " | 1.2023 ±0.0177 | - | 0.0775 |
| | o-e | " | " | " | " | 0.6650 ±0.0285 | - | 0.0656 |
| | o-o | " | " | " | " | 1.1044 ±0.0968 | - | 0.2380 |
| VSm2 | e-e | 1.4833 ±0.0115 | 2.0678 ±1.1487 | -0.2107 ±0.0045 | -30.533 ±0.4533 | - | - | (0.0469) |
| | e-o | 1.5206 ±0.0069 | 4.3178 ±0.2335 | -0.1613 ±0.0006 | -36.2658 ±0.1264 | - | - | 0.0737 |
| | o-e | 1.6077 ±0.0433 | -0.8714 ±3.5349 | -0.1987 ±0.0190 | -34.5795 ±1.5382 | - | - | 0.0552 |
| | o-o | 3.0915 ±0.0103 | -108.543 ±1.6435 | -0.8145 ±0.0056 | 11.3339 ±0.7473 | - | - | 0.1415 |
| SPm1 | e-e | 1.5066 ±0.0015 | -0.2231 ±0.0024 | -29.4184 ±0.1656 | - | - | 0.0 | 0.0473 |
| | e-o | " | " | " | - | - | 0.1609 ±0.0027 | 0.0925 |
| | o-e | " | " | " | - | - | 0.1048 ±0.0053 | 0.0705 |
| | o-o | " | " | " | - | - | 0.2657 ±0.0059 | 0.2504 |
| SPm2 | e-e | 1.5066 ±0.0015 | -0.2231 ±0.0024 | -29.4184 ±0.1656 | - | - | 0.0 | (0.0473) |
| | e-o | 1.549 ±0.0012 | -0.1796 ±0.0020 | -34.095 ±0.2797 | - | - | 0.0441 ±0.0059 | 0.0733 |
| | o-e | 1.5892 ±0.0478 | -0.1897 ±0.0147 | -35.2072 ±0.0318 | - | - | 0.0246 ±0.0993 | 0.0552 |
| | o-o | 10.8468 ±0.6325 | -1.9971 ±0.0924 | -33.9546 ±0.0531 | - | - | -13.4983 ±0.8191 | 0.1378 |

Table 2: The *RMSRE* with its standard deviation *RMS[s(RE)]*, the $\bar{f}$ values, and the *t* statistics for the *2(n-1)* degrees of freedom calculated on the basis of the *RMSRE* and *RMS[s(RE)]*, for the different formulae applied to the TEST nuclides. Included also for comparison are the tabulated *t* values at 90 % and 95% confidence level.

| Formula | Z,N | RMSRE ± RMS[s(RE)] | Avg. RMSRE | $\bar{f}$ | Avg. $\bar{f}$ |
|---|---|---|---|---|---|
| Rm | e-e | 3.812±2.084 | 1.745±0.591 | 2.510 | 4.088 |
|  | e-o | 1.434±1.001 |  | 4.100 |  |
|  | o-e | 0.693±0.484 |  | 3.870 |  |
|  | o-o | 1.039±0.106 |  | 5.873 |  |
| VSm1 | e-e | 3.681±1.671 | 2.247±0.717 | 2.358 | 7.570 |
|  | e-o | 2.593±2.071 |  | 12.00 |  |
|  | o-e | 1.410±0.894 |  | 6.160 |  |
|  | o-o | 1.302±0.594 |  | 9.765 |  |
| VSm2 | e-e | (3.681±1.671) | 1.638±0.456 | (2.358) | 4.440 |
|  | e-o | 1.036±0.545 |  | 3.328 |  |
|  | o-e | 0.753±0.357 |  | 3.990 |  |
|  | o-o | 1.081±0.321 |  | 8.082 |  |
| SPm1 | e-e | 3.763±1.978 | 1.950±0.626 | 2.448 | 5.600 |
|  | e-o | 1.675±1.260 |  | 5.794 |  |
|  | o-e | 1.106±0.682 |  | 4.570 |  |
|  | o-o | 1.257±0.546 |  | 9.589 |  |
| SPm2 | e-e | (3.763±1.978) | 1.775±0.556 | (2.448) | 9.352 |
|  | e-o | 1.106±0.905 |  | 3.588 |  |
|  | o-e | 0.805±0.435 |  | 4.120 |  |
|  | o-o | 1.426±0.167 |  | 27.25 |  |
| Formula pairs | Z,N | t | | $t_{90}$ | $t_{95}$ |

| | | | | |
|---|---|---|---|---|
| VSm2 vs VSm1 | e-o | 5.906 | <1.660 | <1.984 |
| | o-e | 5.057 | <1.660 | <1.984 |
| | o-o | 1.968 | <1.667 | <1.994 |
| SPm2 vs SPm1 | e-o | 2.980 | <1.660 | <1.984 |
| | o-e | 2.762 | <1.660 | <1.984 |
| | o-o | 1.777 | <1.667 | <1.994 |

Table 3: The experimental α decay data, $Q_\alpha^{exp}$ and $T_\alpha^{exp}$, for the EXOTIC nuclides *viz.* the super-heavy elements (SHE) and the medium mass α emitters. The listing from [21] has been upgraded and some recent data [22, 23] included. The results of the calculations using the three modified formulae, the *Rm*, *VSm2* and *SPm2* are presented as $log_{10}T_\alpha^{calc}$ along with their standard deviations. All data have been rounded off to the second decimal place. * Denotes nuclides where a typical 40 keV FWHM experimental error in the decay energy has additionally been incorporated into the calculation to yield the precision error.

### Even-Even Parent

| Parent | Z | A | $Q_\alpha^{exp}$ (MeV) | $T_\alpha^{exp}$ (s) | $Log_{10}T_\alpha^{exp}$ (s) | $Log_{10}T_\alpha^{calc}$ (s) | | |
|---|---|---|---|---|---|---|---|---|
| | | | | | | Rm | VSm2 | SPm2 |
| Er | 68 | 156 | 3.49 | 2.30E10 ±1.2E+08 | 10.36 | 10.42±0.06 | 10.27±0.05 | 10.28±0.06 |
| Yb | 70 | 158 | 4.17 | 4.30E06±3.75E05 | 6.63 | 6.76±0.05 | 6.57±0.03 | 6.60±0.05 |
| W  | 74 | 158 | 6.61±0.00 | 1.50E-03±2.00E-03 | -2.82 | -2.25±0.04 | -2.63±0.00 | -2.57±0.03 |
| Os | 76 | 162 | 6.80±0.00 | 1.90E.03±2.00E-03 | -2.72 | -2.11±0.03 | -2.51±0.00 | -2.45±0.03 |
| Os | 76 | 164 | 6.48 | 4.20E-02±2.00E-03 | -1.38 | -1.07±0.03 | -1.43±0.00 | -1.37±0.03 |
| Pt | 78 | 166 | 7.29 | 3.00E-04 | -3.52 | -2.91±0.03 | -3.34±0.01 | -3.28±0.02 |
| Pt | 78 | 168 | 7.00 | 2.00E-03±4.00E-04 | -2.70 | -2.06±0.03 | -2.44±0.00 | -2.39±0.03 |
| Pt | 78 | 170 | 6.71 | 1.40E-02±1.17E-02 | -1.85 | -1.15±0.03 | -1.50±0.00 | -1.45±0.03 |
| Hg | 80 | 172 | 7.53 | 4.20E-04 | -3.38 | -2.97±0.02 | -3.37±0.01 | -3.33±0.02 |
| Hg | 80 | 174 | 7.23 | 2.10E-03 | -2.68 | -2.13±0.02 | -2.49+0.01 | -2.45±0.02 |
| Pb | 82 | 178 | 7.79 | 2.30E-04 | -3.63 | -3.10±0.02 | -3.49±0.01 | -3.45±0.02 |
| Pb | 82 | 180 | 7.42 | 5.00E-03 | -2.30 | -2.03±0.02 | -2.38±0.01 | -2.34±0.02 |
| Pb | 82 | 184 | 6.77 | 6.10E-01±6.65E-02 | -0.22 | 0.00±0.02 | -0.28±0.00 | -0.25±0.02 |
| Pb | 82 | 186 | 6.47 | 1.20E+01±1.24E-01 | 1.08 | 1.07±0.02 | 0.82±0.00 | 0.86±0.02 |
| Pb | 82 | 194 | 4.74 | 9.80E+09±4.08E+08 | 9.99 | 9.16+0.03 | 9.02±0.01 | 9.05±0.03 |
| Po | 84 | 188 | 8.09±0.025 | 4.00E-04±2.00E-04 -1.50E-04 | -3.40 | -3.40±0.01 | -3.69±0.01 | -3.66±0.01 |
| Po | 84 | 190 | 7.69 | 2.50E-03±6.25E-04 | -2.60 | -2.31±0.01 | -2.56±0.01 | -2.53±0.01 |
| Po | 84 | 192 | 7.32±0.011 | 2.90E-02+1.50E-02 -8.00E-03 | -1.54 | -1.19±0.02 | -1.41±0.00 | -1.38±0.01 |
| Po | 84 | 194 | 6.99±0.003 | 4.215E-01±4.30E-03 | -0.38 | -0.13±0.02 | -0.31±0.00 | -0.28±0.01 |
| Rn | 86 | 196 | 7.62±0.009 | 4.40E-03+1.30E-03 -9.00E-04 | -2.36 | -1.42±0.01 | -1.68±0.00 | -1.65±0.01 |
| Rn | 86 | 198 | 7.35 | 6.50E-02±2.03E-03 | -1.19 | -0.61±0.01 | -0.83±0.00 | -0.81±0.01 |
| Ra | 88 | 202 | 8.02 | 2.60E-03±1.23E-02 | -2.59 | -2.00±0.01 | -2.25±0.01 | -2.23±0.00 |
| Ra | 88 | 204 | 7.64 | 5.90E-02±1.20E-02 | -1.23 | -0.87±0.01 | -1.09±0.00 | -1.07±0.01 |
| Th | 90 | 210 | 8.05±0.02 | 9.00E-03±1.15E-02 | -2.05 | -1.51±0.00 | -1.72±0.01 | -1.71±0.00 |
| U  | 92 | 218 | 8.77±0.01 | 5.10E-04+1.70E-04 - 1.00E-04 | -3.29 | -2.98±-0.01 | -3.14±0.01 | -3.15±0.01 |

| Parent | Z | A | $Q_\alpha^{exp}$ (MeV) | $T_\alpha^{exp}$ (s) | $Log_{10}T_\alpha^{exp}$ (s) | $Log_{10}T_\alpha^{calc}$ (s) | | |
|---|---|---|---|---|---|---|---|---|
| | | | | | | Rm | VSm2 | SPm2 |
| U | 92 | 220 | 10.30 | 6.00E-08 | -7.22 | -6.67±-0.01 | -6.75±0.01 | -6.75±0.01 |
| U | 92 | 224 | 8.62 | 7.00E-04±2.3E-10 | -3.16 | -2.69±-0.01 | -2.73±0.01 | -2.73±0.01 |
| Pu | 94 | 228 | 7.95 | 2.00E-01 | -0.700 | -0.03±-0.01 | -0.15±0.00 | -0.16±0.01 |
| Pu | 94 | 230 | 7.18 | 1.00E+02 | 2.00 | 2.58±0.00 | 2.47±0.00 | 2.46±0.01 |
| Cm | 96 | 238 | 6.62 | 2.30E+05±9.58E+03 | 5.36 | 5.51±-0.01 | 5.39±0.00 | 5.38±0.01 |
| Fm | 100 | 250 | 7.56±0.01 | 2.00E+03±2.00E+02 | 3.30 | 3.24±-0.02 | 3.11±0.00 | 3.08±0.02 |
| No | 102 | 258 | 8.15 | 1.20E+02 | 2.08 | 1.80±-0.02 | 1.70±0.00 | 1.65±0.02 |
| Rf | 104 | 256 | 8.95 | 3.04E-01 | -0.52 | 0.07±-0.03 | -0.19±0.00 | -0.25±0.03 |
| Rf | 104 | 258 | 9.25 | 9.20E-02±1.53E-02 | -1.04 | -0.83±-0.03 | -1.04±0.00 | -1.10±0.03 |
| Rf | 104 | 260 | 8.901 | 1.00E+00±3.48E-02 | 0.00 | 0.14±-0.03 | -0.04±0.00 | -0.10±0.03 |
| Sg | 106 | 260 | 9.92 | 7.20E-03 | -2.14 | -1.99±-0.03 | -2.29±0.01 | -2.36±0.04 |
| Hs | 108 | 264 | 10.8 | 8.1E-05 | -4.09 | -3.56±-0.04 | -3.91±0.01 | -4.00±0.04 |
| Hs | 108 | 266 | 10.34 | 2.30E-03 | -2.64 | -2.50±-0.04 | -2.82±0.01 | -2.91±0.04 |
| Hs | 108 | 270 | 9.02 | 2.20E+01 | 1.34 | 1.04±-0.03 | 0.74±0.01 | 0.67±0.04 |
| Ds | 110 | 270 | 11.20 | 1.00E-04 | -4.00 | -3.96±-0.04 | -4.33±0.01 | -4.44±0.05 |
| Ds* | | | | | | (-3.96±0.13)* | (-4.33±0.08)* | (-4.44±0.14)* |
| Uuq | 114 | 286 | 10.35±0.06 | 1.60E-01+7.00E-02 -3.00E-02 | -0.80 | -0.94±-0.05 | -1.34±0.01 | -1.46±0.05 |
| Uuq | 114 | 288 | 10.09±0.07 | 8.00E-01+3.20E-01 -1.80E-01 | -0.10 | -0.28±-0.05 | -0.66±0.01 | -0.78±0.05 |
| Uuh | 116 | 290 | 11.00±0.08 | 1.50E-02+2.60E-02 -6.00E-03 | -1.82 | -2.00±-0.05 | -2.47±0.02 | -2.60±0.06 |
| Uuh | 116 | 292 | 10.80±0.07 | 1.80E-02+1.60E-02 -6.00E-03 | -1.75 | -1.55±-0.05 | -1.98±0.02 | -2.12±0.06 |
| Uuo | 118 | 294 | 11.81±0.06 | 1.80E-03+7.50E-02 -1.30E-03 | -2.75 | -3.33±-0.06 | -3.86±0.02 | -4.01±0.06 |

Even-Odd Parent

| Parent | Z | A | $Q_\alpha^{exp}$ (MeV) | $T_\alpha^{exp}$ (s) | $Log_{10}T_\alpha^{exp}$ (s) | $Log_{10}T_\alpha^{calc}$ (s) | | |
|---|---|---|---|---|---|---|---|---|
| | | | | | | Rm | VSm2 | SPm2 |
| Te | 52 | 105 | 4.90 | 7.00E-07+2.5E-07 -1.7E-07 | -6.15 | -6.35±0.12 | -6.98±0.10 | -6.88±0.13 |
| Po | 84 | 189 | 7.70±0.02 | 5.00E-03±1.0E-03 | -2.30 | -1.95±0.06 | -2.24±0.06 | -2.17±0.06 |
| Po | 84 | 209 | 4.98±0.00 | 3.23E+09±1.59E+08 | 9.51 | 9.34±0.03 | 9.36±0.03 | 9.39±0.03 |
| Ra | 88 | 223 | 5.98±0.00 | 9.88E+05±3.46E+05 | 5.99 | 5.95±0.03 | 6.03±0.03 | 6.05±0.04 |
| Th | 90 | 227 | 6.15±0.00 | 1.62E+06±1.73E+05 | 6.21 | 6.12±0.03 | 6.16±0.03 | 6.18±0.03 |
| Fm | 100 | 245 | 8.44 | 4.20E+00±1.30E+00 | 0.62 | 1.55±0.03 | 1.43±0.03 | 1.40±0.02 |
| Rf | 104 | 259 | 9.12 | 3.33E+00±7.66E-01 | 0.52 | 0.84±0.03 | 0.75±0.03 | 0.70±0.02 |
| Sg | 106 | 259 | 9.83±0.03 | 5.33E-01±2.35E-01 | -0.27 | -0.41±0.03 | -0.59±0.04 | -0.66±0.01 |
| Sg | 106 | 271 | 8.65±0.08 | 1.44E+02±2.58E+02 -6.00E+01 | 2.16 | 2.93±0.02 | 2.91±0.03 | 2.84±0.01 |
| Hs | 108 | 275 | 9.44±0.07 | 1.50E-01+2.70E-01 -6.00E-02 | -0.82 | 1.21±0.02 | 1.17±0.03 | 1.08±0.01 |
| Hs | 108 | 267 | 10.11 | 2.60E-02±1.50E-02 | -1.59 | -0.54±0.03 | -0.69±0.04 | -0.77±0.01 |
| Hs* | | | | | | (-0.54±0.08)* | (-0.69±0.07)* | (0.77±0.10)* |

| Parent | Z | A | $Q_\alpha^{exp}$ (MeV) | $T_\alpha^{exp}$ (s) | $Log_{10}T_\alpha^{exp}$ (s) | $Log_{10}T_\alpha^{calc}$ (s) | | |
|---|---|---|---|---|---|---|---|---|
| | | | | | | Rm | VSm2 | SPm2 |
| Ds | 110 | 279 | 9.84±0.06 | 1.80E-01+5.00E-02 -3.00E-02 | -0.74 | 0.77±0.02 | 0.69±0.03 | 0.59±0.00 |
| Uub | 112 | 283 | 9.67±0.06 | 4.00E+00+1.30E+00 -7.00E-01 | 0.60 | 1.96±0.01 | 1.82±0.03 | 1.71±0.00 |
| Uub | 112 | 285 | 9.29±0.06 | 3.40E+01+1.70E+01 -9.00E+00 | 1.53 | 3.09±0.01 | 2.96±0.02 | 2.85±0.00 |
| Uuq | 114 | 287 | 10.16±0.06 | 5.10E-01+1.80E-01 -1.00E-01 | -0.29 | 1.26±0.01 | 1.08±0.03 | 0.95±0.00 |
| Uuq | 114 | 289 | 9.96±0.06 | 2.70E+00+1.4E+00 -0.7E+00 | 0.43 | 1.79+0.01 | 1.64±0.03 | 1.51±0.01 |
| Uuh | 116 | 291 | 10.89±0.07 | 6.30E-03+1.16E-02 -2.50E-03 | -2.20 | -0.01±0.02 | -0.22±0.03 | -0.37±0.01 |
| Uuh | 116 | 293 | 10.67±0.06 | 5.3E-02+6.2E-02 -1.9E-02 | -1.28 | 0.52±0.01 | 0.34±0.03 | 0.19±0.01 |

Odd-Even Parent

| Parent | Z | A | $Q_\alpha^{exp}$ (MeV) | $T_\alpha^{exp}$ (s) | $Log_{10}T_\alpha^{exp}$ (s) | $Log_{10}T_\alpha^{calc}$ (s) | | |
|---|---|---|---|---|---|---|---|---|
| | | | | | | Rm | VSm2 | SPm2 |
| Ir | 77 | 175 | 5.71±0.00 | 1.06E+03±4.21E+02 | 3.02 | 1.61±0.10 | 1.56±0.00 | 1.51±0.01 |
| Au | 79 | 175 | 6.68 | 2.13E-01±4.22E-02 | -0.67 | -1.35±0.11 | -1.48±0.00 | -1.53±0.03 |
| Np | 93 | 231 | 6.37±0.05 | 1.46E+05±7.35E+04 | 5.17 | 5.89±0.05 | 5.84±0.03 | 5.83±0.03 |
| Am | 95 | 237 | 6.18±0.01 | 1.75E+07±3.19E+06 | 7.24 | 7.72±0.04 | 7.62±0.03 | 7.62±0.03 |
| Es | 99 | 241 | 8.32 | 9.00E+00 | 0.95 | 0.83±0.06 | 0.62±0.08 | 0.64±0.11 |
| Md | 101 | 249 | 8.46 | 1.20E+02±6.33E+01 | 2.08 | 1.08±0.05 | 0.87±0.09 | 0.90±0.12 |
| Lr | 103 | 255 | 8.61 | 2.59E+01±4.86E+00 | 1.41 | 1.32±0.05 | 1.07±0.10 | 1.11±0.13 |
| Db | 105 | 263 | 8.83 | 6.28E+01±2.95E+01 | 1.80 | 1.32±0.04 | 1.07±0.11 | 1.11±0.13 |
| Db* | | | | | | (1.32±0.09)* | (1.07+0.13)* | (1.11+0.15)* |
| Bh | 107 | 261 | 10.56±0.05 | 1.24E-02±4.26E-02 | -1.91 | -2.80±0.06 | -3.18±0.15 | -3.12±0.21 |
| Mt | 109 | 275 | 10.48±0.09 | 9.70E-03±4.60E-02 -4.40E-03 | -2.01 | -2.08±0.05 | -2.38±0.16 | -2.31±0.22 |
| Rg | 111 | 279 | 10.52±0.20 | 1.70E-01+8.10E-01 -8.00E-02 | -0.77 | -1.50±0.04 | -1.89±0.17 | -1.81±0.22 |
| Uut | 113 | 283 | 10.26±0.09 | 1.00E-01+4.90E-01 -4.50E-02 | -1.00 | -0.11±0.03 | -0.59±0.18 | -0.51±0.22 |
| Uup | 115 | 287 | 10.74±0.09 | 3.20E-02+1.55E-01 -1.40E-02 | -1.49 | -0.72±0.03 | -1.29±0.20 | -1.19±0.24 |

Odd-Odd Parent

| Parent | Z | A | $Q_\alpha^{exp}$ (MeV) | $T_\alpha^{exp}$ (s) | $Log_{10}T_\alpha^{exp}$ (s) | $Log_{10}T_\alpha^{calc}$ (s) | | |
|---|---|---|---|---|---|---|---|---|
| | | | | | | Rm | VSm2 | SPm2 |
| Re | 75 | 162 | 6.27±0.05 | 1.14E-01±1.56E-02 | -0.94 | -1.02±0.03 | -0.52±0.02 | -0.78±0.02 |
| Tl | 81 | 184 | 6.30±0.05 | 5.24E+02±1.81E+02 | 2.72 | 1.56±0.03 | 1.89±0.02 | 1.75±0.03 |
| Np | 93 | 226 | 8.20±0.05 | 3.5E-02 | -1.46 | -1.23±0.00 | -1.91±0.01 | -3.11±0.03 |
| Am | 95 | 234 | 6.87 | 3.57E+05±1.11E+05 | 5.55 | 5.11±0.03 | 4.59±0.03 | 4.65±0.03 |
| Es | 99 | 246 | 7.74 | 4.67E+03±9.01E+02 | 3.67 | 2.94±0.01 | 1.69±0.03 | 1.35±0.03 |
| Md | 101 | 248 | 8.70 | 3.50E+01±2.30E+01 | 1.54 | 0.10±0.01 | -1.87±0.02 | -3.14±0.04 |
| Lr | 103 | 254 | 8.85 | 1.67E+01±2.87E+00 | 1.22 | 0.32±0.01 | -2.01±0.02 | -3.32±0.04 |
| Db | 105 | 258 | 9.55 | 6.57E00±1.31E+00 | 0.82 | -1.27±0.02 | -4.26±0.02 | -6.39±0.06 |
| Bh | 107 | 264 | 9.97 | 4.40E-01 | -0.36 | -1.92±0.03 | -5.43±0.02 | -8.06±0.07 |
| Bh | 107 | 272 | 9.15±0.06 | 9.80+00+1.17E01 -3.50E+00 | 0.99 | 0.62±0.02 | -2.35±0.02 | -3.76±0.05 |
| Mt | 109 | 276 | 9.85+0.06 | 7.2E-01+8.7E-01 -2.5E-01 | -0.14 | -0.97±0.03 | -4.66±0.02 | -6.95±0.07 |
| Rg Rg* | 111 | 272 | 10.98 | 1.51E-03 | -2.82 | -3.46±0.04 (-3.46±0.15)* | -8.27±0.02 (-8.27±0.11)* | -12.28±0.11 (-12.28±0.20)* |
| Rg | 111 | 280 | 9.87±0.06 | 3.6E+00+4.3E+00 -1.3E+00 | 0.56 | -0.29±0.03 | -4.40±0.03 | -6.56±0.08 |
| Uut | 113 | 282 | 10.63±0.08 | 7.30E-02+0.13 -0.03 | -1.14 | -1.89±0.04 | -6.85±0.03 | -10.09±0.10 |
| Uut | 113 | 284 | 10.15+0.06 | 4.8E-01+0.58 -0.17 | -0.32 | -0.46±0.03 | -5.12±0.03 | -7.58±0.09 |
| Uup | 115 | 288 | 10.61±0.06 | 8.70E-02+0.11 -0.03 | -1.06 | -1.17±0.04 | -6.51±0.03 | -9.57±0.11 |

Table 4: The *RMSRE* and its average value for the different formulae and their linearly optimized variants for the odd parity sets, as obtained for the EXOTIC nuclei.

| Formula | Z,N | RMSRE* | Avg. RMSRE (all parities) | Avg. RMSRE (odd parities) |
|---|---|---|---|---|
| R | e-e | 0.391 | 0.761 | 0.884 |
|   | e-o | 1.137 |  |  |
|   | o-e | 0.573 |  |  |
|   | o-o | 0.943 |  |  |
| Rm | e-e | 0.448 | 1.209 | 1.462 |
|   | e-o | 1.841 |  |  |
|   | o-e | 0.539 |  |  |
|   | o-o | 2.006 |  |  |
| Lin-opt-Rm | e-o | 1.518 | - | 0.994 |
|   | o-e | 0.502 |  |  |
|   | o-o | 0.961 |  |  |
| VSS | e-e | 1.018 | 1.047 | 1.056 |
|   | e-o | 1.468 |  |  |
|   | o-e | 0.589 |  |  |
|   | o-o | 1.113 |  |  |
| VSm2 | e-e | 0.915 | 3.330 | 4.134 |
|   | e-o | 1.674 |  |  |
|   | o-e | 0.630 |  |  |
|   | o-o | 10.099 |  |  |
| Lin-opt-VSm2 | e-o | 1.485 | - | 3.378 |
|   | o-e | 0.509 |  |  |
|   | o-o | 8.141 |  |  |
| SP | e-e | 1.108 | 1.038 | 1.014 |
| | e-o | 1.296 | | |
| | o-e | 0.693 | | |
| | o-o | 1.054 | | |
| SPm2 | e-e | 1.094 | 4.613 | 5.786 |
| | e-o | 1.547 | | |
| | o-e | 0.624 | | |
| | o-o | 15.187 | | |
| Lin-opt-SPm2 | e-o | 1.331 | - | 3.550 |
| | o-e | 0.503 | | |
| | o-o | 8.816 | | |

\* No calculation of the net standard deviation RMS[s(RE)] done because of insufficient information on the experimental errors (*vide* table 3) and coefficient errors of the original R, VSS and SP formulae. It is not essential for the discussion relevant to this table.

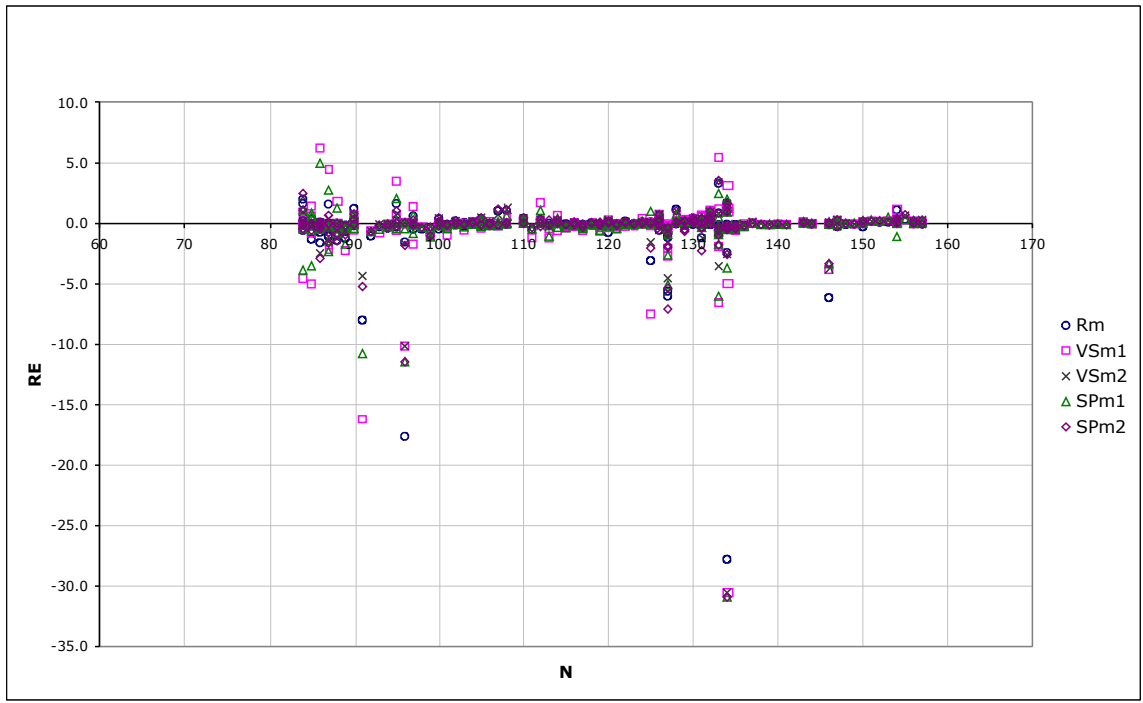

Fig. 1: Plot of the variation of the relative error (RE) between the $\log_{10}$ of the calculated and experimental $\alpha$ half-lives (eqn. 4) with the neutron number, N, for the TEST nuclei.

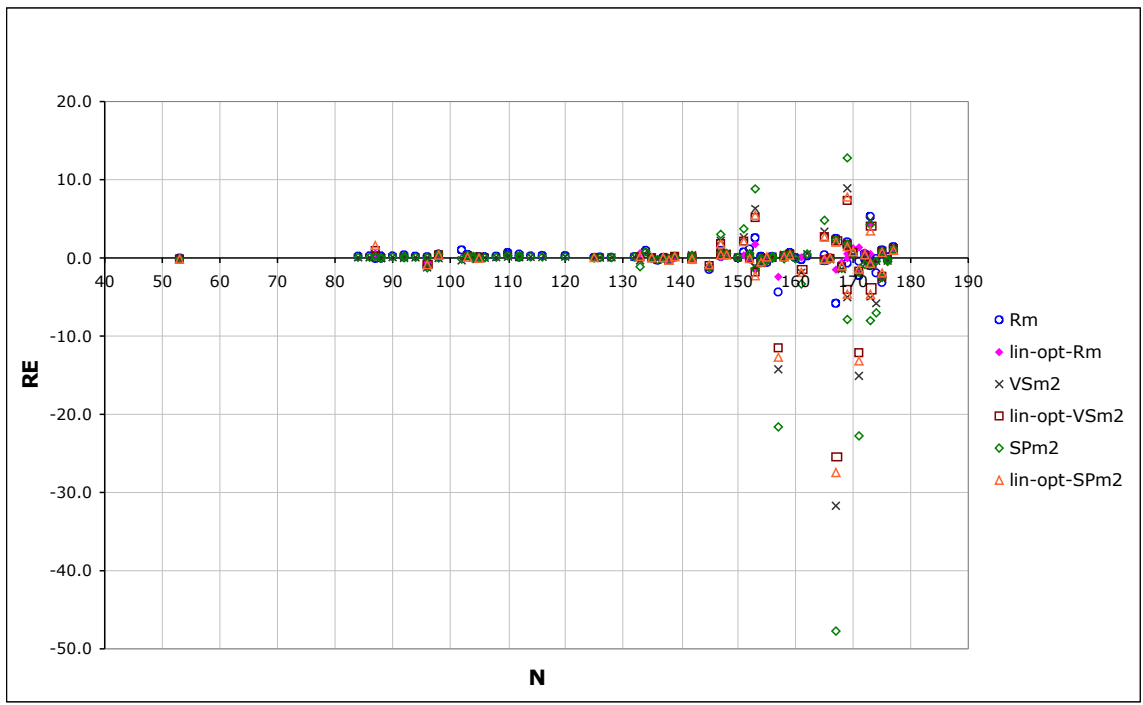

Fig. 2: Plot of the variation of the relative error (RE) between the $\log_{10}$ of the calculated and experimental $\alpha$ half-lives with the neutron number, N, for the EXOTIC nuclei.